\documentclass{svproc}
\usepackage{graphicx}
\usepackage[misc]{ifsym}
\usepackage{url}

\begin{document}
\mainmatter              
\title{A Transformer Based Pitch Sequence Autoencoder with MIDI Augmentation}
\titlerunning{A Transformer Based Autoencoder}  
%
\author{Mingshuo Ding\inst{1} \and Yinghao Ma\inst{2}\Letter}
\authorrunning{Mingshuo Ding et al.} 
%
\tocauthor{Mingshuo Ding, Yinghao Ma}
\institute{Peking University, Beijing 100871, CHN,\\
\email{dingmingshuo@pku.edu.cn}
\and
Peking University, Beijing 100871, CHN,
\\
\email{\email{yhma625@pku.edu.cn}}}

\maketitle              

\begin{abstract}
Despite recent achievements of deep learning automatic music generation algorithms, few approaches have been proposed to evaluate whether a single-track music excerpt is composed by automatons or Homo sapiens. 
To tackle this problem, we apply a masked language model based on ALBERT for composers classification. 
The aim is to obtain a model that can suggest the probability a MIDI clip might be composed condition on the auto-generation hypothesis, and which is trained with only AI-composed single-track MIDI.
In this paper, the amount of parameters is reduced, two methods on data augmentation are proposed as well as a refined loss function to prevent overfitting. 
The experiment results show our model ranks $3^{rd}$ in all the $7$ teams in the data challenge in CSMT(2020). 
Furthermore, this inspiring method could be spread to other music information retrieval tasks that are based on a small dataset.
\end{abstract}
\section{Introduction}
Methods based on machine learning have been widely proposed for automatic music generation since significant progress on deep learning. 
Nowadays, more and more melodies can be composed by artificial intelligence through using the pitch and length of the notes in human music as primary inputs to mimic humans \cite{liu2016computational,dong2017musegan,li2020comparison}. 
Unlike checking counterpoint in multi-track melodies and evaluation self-similarity matrix in music structure analysis, few objective algorithms or indicators have been put forward to assess whether a single-track short melody is created by a machine or a person. 
Although several attempts has been made, such as measures from information theory to compare Bach’s music \cite{ren2015using}, or probability transfer relation with the N-gram model to compare British and American folk music melody \cite{li2020comparison}, most of the classification model on composers works are based on human opinions, namely, the participants listened to a music excerpt and then judged whether it was composed by a human or an AI \cite{liang2017automatic,chu2016song,huang2016deep,unehara2001composition}.

However, the result of listening tests might contain individual or group differences, which makes them difficult to be compared among different people, especially when the amount of samples is small.
Finding a relatively common and objective approach to classify the composer of a short piece of melodies in various musical styles can make different music generation models comparable. 
The purpose of this study is to find an objective and effective method to generate an indicating value of whether a music clip is human-composed by analyzing the AI-made melodies.

Features extracting is an essential component for music series related tasks. For the single-track data without chords, there are some methods that rely on N-gram\cite{li2020comparison,ogihara2008n}. However, this approach is difficult to model the long-term dependence and the following dependence, and the data is sparse with the exponential growth of probability as sequence length increases, which leads to poor generalization ability. Besides, Bidirectional Encoder Representations from Transformers (BERT, Fig.1) \cite{devlin2018bert} might be a promising technique except its large amount of parameters such as learning an embedding for a sequence after parameters reduction \cite{li2020comparison}.
	
\begin{figure}
\centering
\includegraphics[width=180 pt]{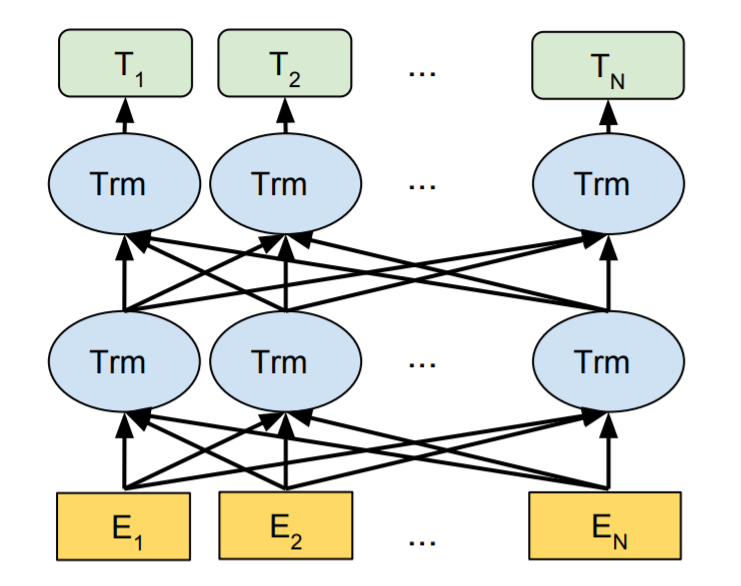}
\caption{BERT uses a bidirectional Transformer.\cite{devlin2018bert}}
\end{figure}

In fact, BERT as a pre-trained models \cite{devlin2018bert,peters2018deep,radford2018improving} has dominated the field of Natural Language Processing (NLP) in the past two years. This model uses self-supervised learning to encode contextual information to obtain a powerful and universal representation. 
This representation can improve performance, especially in situations where data for downstream tasks is limited. More recently, BERT-like models have been applied to speech processing \cite{liu2020mockingjay,jiang2019improving,ling2020deep,schneider2019wav2vec}. 
However, such models usually maintain a large number of parameters in both speech tasks and text tasks, requiring a large amount of data and memory for training and computation. Therefore, it might be prone to overfit when pre-training data is relatively scarce, such as in music related cases.

A Lite BERT (ALBERT) \cite{lan2019albert} is a simplified version of BERT that shares the same parameters at all layers and decompose the embedding matrix to reduce most of the parameters. Although the number of parameters is reduced, the representation learned in ALBERT is still robust and task agnostic, so that ALBERT can achieve similar performance to BERT in the same downstream task \cite{chi2020audio}, thus is also regarded as obtaining characteristics about the input itself.

In this paper, a masked language model (MLM) which is based on ALBERT is introduced into MIDI processing and a new self-supervised model is proposed.

The rest of this article is organized as follows. In section 2, the dataset used in the study is described as well as the data preprocessing and strategies used for data augmentation. In section 3, the pipeline of the research, the methods on prevention of overfitting are demonstrated, as well as the detail of the ALBERT model and the approach to evaluate the probability of each composer. Section 4 covers the main experimental processes and results. The fifth section we have made the summary and the prospect.

\section{Dataset}
\subsection{Training Data}
The data set is provided in the data challenge of Conference of Sound and Music Technology (CSMT) 2020 \cite{li2020csmt}. The training data only contains the music generated by artificial intelligence algorithms which includes 6000 MIDI files. Each file is single melodic music whose speed is between 68BPM and 118BPM. Each melody is 8-bar length, without complete phrase structure. In fact, complete music sentences are always with 8 or 16 bars and this suggests that the start point of each music excerpt is not the beginning of any music sentences. Besides, it should be noted that the melodies in the training data set are generated by several machine models trained with data in two unannounced different music genres. More information can be found at the website \footnote{http://www.csmcw-csmt.cn/data/2020/ai-composition-recognition2020/?from=timeline}.

Despite many open source MIDI datasets on the internet such as the one on reddit with 3.65 GB multi-track MIDI in all sorts of music genre\footnote{https://www.reddit.com/r/datasets/comments/3akhxy}, the single-track music clips like what is provided in the data challenge are rare, not to mention the uncertainty on music genre. As a consequence, it is difficult to extract a convincing main melody especially condition on similar music range and notes distribution. Therefore, training did NOT use any human composed data.

\subsection{Data Preprocessing}
For the specific problem of comparing the similarities of melodies, the rhythm and pitch are important characteristics, since people usually pay attention to them when they perceive music melodies \cite{kim2000analysis}. Thus, the MIDI sequence of 8 bars can be segmented into 128 hexadecimal notes or 256 thirty-second notes, as the speed and the starting and ending time of the notes are marked.  Whether the unit of the 8-bar music is a hexadecimal note or a thirty-second note depends on the shortest note length in the given MIDI, and there are 256 notes or so in a music sequence for most of the cases. Considering the fact that it is meaningless in music to divide a quarter note into twelve equal parts in the vast majority of cases, there is no musical necessity to do so except for compatibility with the relative rarity of triplets and sixteenth notes. Thus, we classify all triplets as three quavers or three sixteenth notes in the same probability, which leads to the total length of a music sequence not being 256. Given that the speed of each music piece is uniformed as the tempo of each music piece is similar to Andrate, the feature of speed in each MIDI sequence is not taken into consideration. In this way, each single-track MIDI clip is turned into a pitch sequence.

\subsection{Data Augmentation}
Although a noticeable amount of parameters has been decreased in ALBERT relative to BERT parameters, 6000 MIDI data are somehow relatively poor for training. As a consequence, it is vital to adopt some measures on data augmentation. Unfortunately, data augmentation methods usually used in NLP tasks \cite{wei2019eda} can be seldom used in music series processing. 

Randomly swapping is a common approach, but the exchange of music notes may cause non-negligible differences in feeling for a human listener. Music clips for the composition of humanity, for example several sixteenth notes in a crotchet or half note exchange with other sounds, could lead to a strange auditory experience, and let the audience regard the music piece as machine-created. Synonym replacement is not suitable in a sequence of music analysis, because there is no specific semantic like natural language for music notes or sequences. Therefore, it's hard to define whether two notes are “synonym”. Even replacing the octave “synonym” is unacceptable in a lyrical semiquaver with a long note, which results in a clear change in music expressed in human emotion, though little differences infrequency spectrum. In addition, Random insert and delete run a high risk which could make melody strange and weird. It is also hard to change the music from major mode to minor mode for augmentation because the mode is hard to find with only single-track especially without music sentences in it. Moreover, the tempo change augmentation can be hardly used either as the tempo is already uniformed. So we proposed two methods to augment data. 

\begin{figure}
\centering
\includegraphics[width=.6\textwidth]{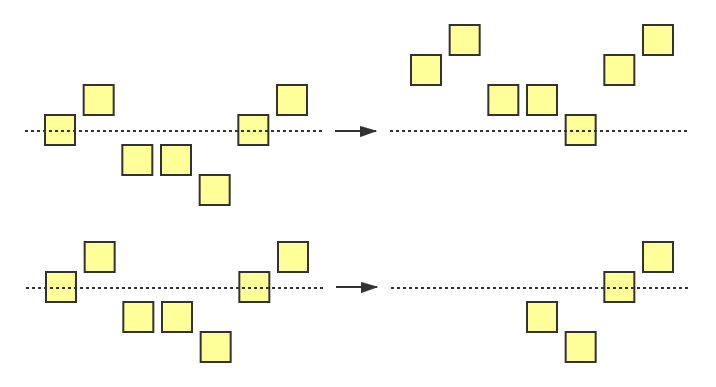}
\caption{Data augmentation approaches: transposition and random truncation}
\end{figure}

\subsubsection{Transposition}
The first data augmentation measure taken in our research is transposition in music tunes. Since music does not make a significant difference, at least not in the respect whether it is generated by human beings or artificial intelligence if it is just changed in music mode. 

Each time, a transposition raises or lowers all the notes in the same pitch sequence by a same random music interval. All the positions the MIDI clips might be transposed to is restricted by both the MIDI range $128$ and the music range, that is the highest note subtract the lowest note. The number of cases for a certain music piece $num$ is as follows, including zero transposition:

\begin{equation}
  num = 128 - highest + lowest + 1.
\end{equation}

Each MIDI transposition is implemented with the same possibility to all the cases. In this way, several relatively same melodies in different music tunes are generated by the transposition data augmentation.
\subsubsection{Random Truncation}
In addition, BERT's training results contain position embedding and thus absolute position information \cite{lee2019set}, for example the word at the beginning of the sentence may be regarded as the subject of the sentence. But the dataset neither includes complete phrase information nor cadence in multi-track, therefore, some location information in the training set retained by BERT belongs to some kind of over-fitting. In order to give up this information, we randomly delete the first few notes of each pitch sequence for the model.

\section{Methods}
The pipeline of our model is shown in Fig.2. First of all, the training set will undergo a data preprocessing part as described above and be expanded by the two data augmentation approaches.Secondly, a MLM task based on ALBERT is trained with refined loss function for an autoencoder on the expanded training set. Lastly, the trained model will be used for evaluation.

\begin{figure}[h]
\centering
\includegraphics[width=.65\textwidth]{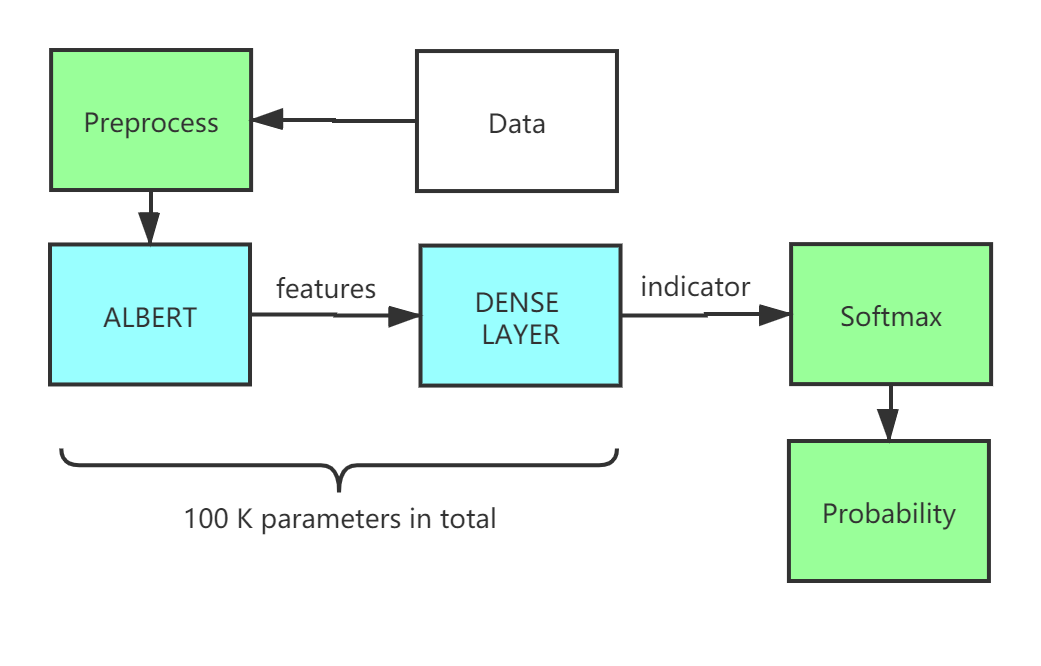}
\caption{Flow Chart of Data Processing}
\end{figure}

\subsection{Avoid Over-fitting}
Since there is only machine-generated data used and no data on human composition, it is still easy to overfit even after data augmentation. To cope with this problem, several additional measures have been taken to prevent from data overfitting. 
\subsubsection{Refined Loss Function}
Some studies have shown that slight adjustment of the loss function $l$ can prevent overfitting greatly \cite{ishida2020we}:

\begin{equation}
l_{new} = \left| l_{origin} - b\right| + b,
\end{equation}

where $b$ is a little positive real parameter which is problem related. The model is trained with the refined loss function and $b$ is set to $0.05$ which is a magic number in some NLP tasks to prevent from pursuing zero-value of original loss function but only to a close-zero value.
\subsubsection{Smaller Transformer}
The number of parameters in the BERT model is extremely large. Even in the ALBERT model using shared parameters, the number of parameters can easily lead to overfitting on such a small dataset. Therefore, on the basis of retaining the structure of ALBERT, the dimension of embedding is $64$, the number of multi-layers is set to $2$ as well as the number of multi-head is $4$. As a result, the amount of parameters of ALBERT is reduced significantly to around 103.6k, thus avoiding potentially overfitting on the training set.
\subsection{Training Method}
There are two important tasks of Bert's training process \cite{devlin2018bert}: Masked Language Model (MLM) and Next Sentence Prediction (NSP). However, the NSP task is not necessary in this problem, because the training set does not include complete phrase information. Actually, it will be hard to divide notes into several phrases. On the contrary, MLM is suitable to tackle this problem.

We hope that the AI composing algorithms used in the dataset which is relatively certain can be fitted through the coding representation obtained by the more "universal" ALBERT with a large number of parameters. Some items of the MIDI sequence is masked and predictions are made on each of the masked note position based on the corresponding embedding vector learned by ALBERT. Such predictions might be closer to the results of some of the AI composers than to those of humans. Assuming that the music was composed by an algorithm fitted by the ALBERT model, the average ``probability'' of each masked note being the same as its ground truth note can be seen as the ``P-value'' indicating whether it was created by AI and the hypothesis shall be accept or reject.

Note that ALBERT training will randomly mask N-grams to make predictions \cite{lan2019albert}. If the masking happens to cover a whole bar or a whole chord formed by adjacent notes, the notes masked are difficult to be effectively predicted.

After comprehensive consideration, the MLM task is the only used task for training. Each time, about 15\% of the elements has been randomly masked in a pitch sequence, and then use the other elements not masked to predict the elements that have been masked. Selecting 15\% notes can ensure that the essential music components are not masked, so that the model can produce effective prediction, and random selection can avoid overfitting to a certain extent as well. And the softmax cross-entropy is used as the loss function of the model to evaluate the distance between the one-hot vector ground truth and the 128 dimensions vector representing the probability of being each of the 128 MIDI notes, followed by the process mentioned above to refine the loss.
\subsection{Evaluation}
When evaluating, for a pitch sequence, each note will be masked successively. Then, the probability $p_i$ of the $i^{th}$ masked note is predicted by the trained ALBERT, and the average probability of all notes is the probability that this data is composed by AI. Formally, the number of notes in this pitch sequence is denoted as $n$, and suggests the probability of AI generating is as follows:
\begin{equation}
p=\frac{1}{n}\sum_{i=1}^{n}{p_i}
\end{equation}
Thus, the probability of each data created by humans, which this task required, can be obtained by $1-p$.
\section{Experiment}
\subsection{Data Setup}
Based on the Albert model, the autoencoder model is trained with MLM tasks on the dataset provided by CSMT(2020) after augmenting.
Both data augmentation strategies mentioned above are used for all the data in the training set.

Firstly, we use \textit{pretty\_midi} \cite{raffel2014intuitive} reads the data in and then preprocesses it. For a pitch sequence after preprocessing, 31 different transpositions are generated including the case remaining the same. And16 of them are implemented with different values of random truncation range in 1to 100. Due to the fact that there are only 12 different modes in an octave and the limitation of computing resources, the size of the augmentation is not extremely large and only part of them are used for training.Therefore, the size of the training set is expanded to $186000$, which is enough for training on the small ALBERT.
\subsection{Environment and Hyper parameters}
Under the good parameter control strategy, the Albert is able to be deployed on a GTX 1050Ti NVIDIA graphic card. \textit{Pytorch} \cite{paszke2019pytorch} and \textit{Hugging Face} \cite{Wolf2019HuggingFacesTS} are used in the process of building and training the algorithm. The small batch size is 64 and the default learning rate is $10^{-3}$ with AdamW optimizer \cite{loshchilov2017decoupled}. The parameter $b$ mentioned is set as 0.05. Because there is no ground truth in the test set, we can not carry out the ablation experiment, the selection of hyper parameters is all based on past experience.

\subsection{Experiment Result}
The data challenge uses the average under receiver operating characteristic curve (AUC) as an indicator for each model performance.The overall performance of AUC is 0.6821 which is rank $4^{th}$ in the 9 models including the baseline model and rank $3^{rd}$ in all of the 7 teams that finished the data challenge. 

The details of the result are shown in the following table.

\begin{table}
\caption{The AUC of test data in different music style}
\begin{center}
\begin{tabular}{r@{\quad}rl}
\hline
\multicolumn{1}{l}{\rule{0pt}{12pt}
                   style}&\multicolumn{2}{l}{AUC value}\\[2pt]
\hline\rule{0pt}{12pt}
J.S.Bach  &     0.6984 & \\
pop song  &    0.6673& \\[2pt]
\hline
\end{tabular}
\end{center}
\end{table}

\begin{table}
\caption{The AUC of test data composed by different AI algorithm}
\begin{center}
\begin{tabular}{r@{\quad}rl}
\hline
\multicolumn{1}{l}{\rule{0pt}{12pt}
                   algorithms}&\multicolumn{2}{l}{AUC value}\\[2pt]
\hline\rule{0pt}{12pt}
GAN  &     0.7458& \\
Transformer&    0.7811& \\
VAE  &    0.3210& \\[2pt]
\hline
\end{tabular}
\end{center}
\end{table}
\begin{table}
\caption{The AUC of test data composed by human}
\begin{center}
\begin{tabular}{r@{\quad}rl}
\hline
\multicolumn{1}{l}{\rule{0pt}{12pt}
                   category}&\multicolumn{2}{l}{AUC value}\\[2pt]
\hline\rule{0pt}{12pt}
published  &     0.6895& \\
unpublished &    0.5404& \\[2pt]
\hline
\end{tabular}
\end{center}
\end{table}
The AUC values of different music styles do not show significant difference, which implies our model may keep an objective evaluation among different music styles. Furthermore, the result of VAE composed is extremely low, even worse than the random guess. Although the test data is not published and audios can not be listened for finding some missing patterns, this phenomenon deserves more attention. Finally, the unpublished result is a bit lower than the published data. This might be caused by the relatively small number of unpublished data and these data are composed by conservatory students instead of composers like Bach and these might keep some difference with each other.

\section{Conclusion}
In this paper, we proposed an autoencoder approach based on ALBERT with the aim to set up an indicator to reject the hypothesis that the music excerpt is composed by machine. The ALBERT model is trained self-supervised with a MLM to mimic the AI-composer. Experimental results confirmed that the brand-new method outperforms some of other algorithms and rank $3^{rd}$ and shows little difference in two music styles. Besides, we found the model performance on VAE models is extremely low, therefore, deserve more attention.

Our model provides a meaningful approach and can be spread to similar tasks with small dataset. However, there are several problems unavoidable as well. To begin with, the whole semantics of the encoder is hard to be understood as the performance on some of the models is relatively high and others are extremely low, which suggest the obvious uncertainty on there liability of the workflow. In addition, the indicator in our model based on the encoder works in the way of p-value and keeps some weakness by nature. Some good music pieces may have high probability to be composed by both human composers and artificial intelligence and other weird MIDI clips might be low possibility to be composed by both homo sapiens and automatons. These unsolid pseudo p-values shall be avoided or be implemented in great caution when it is spread to other tasks if there are some data in another class.

\bibliographystyle{spphys}
\bibliography{mybib}
\end{document}